\documentclass{article}

   \usepackage[preprint]{neurips_2024}

\setcitestyle{numbers,square,comma}

\usepackage[utf8]{inputenc} 
\usepackage[T1]{fontenc}    
\usepackage{hyperref}       
\usepackage{url}            
\usepackage{booktabs}       
\usepackage{amsfonts}       
\usepackage{nicefrac}       
\usepackage{microtype}      
\usepackage{xcolor}         
\usepackage{graphicx}
\usepackage{subcaption}
\usepackage{caption}
\usepackage{float}
\usepackage[justification=raggedright]{caption}

\title{Intraoperative Glioma Segmentation with YOLO + SAM for Improved Accuracy in Tumor Resection}

\author{
  Samir Kassam\thanks{Lead author.}\\
  Carnegie Vanguard High School\\
  {samir.s.kassam@gmail.com} \\
\And
 Angelo Markham\textsuperscript{*}\\
 Devon Preparatory High School  \\
 amarkham@devonprepstudents.org\\
\And
 Katie Vo\textsuperscript{}\\
 Lake Travis High School  \\
 katievo2017@gmail.com \\
 \And
 Yashas Revanakara\\
 Irvington High School  \\
 yashasr471@gmail.com \\
 \AND
 Michael Lam\textsuperscript{\ddag}\\
 Algoverse AI Research  \\
 michael@algoverse.us \\
 \And
 Kevin Zhu\textsuperscript{\ddag} \\
 Algoverse AI Research  \\
 kevin@algoverse.us \\
 }

\begin{document}

\maketitle
\vspace{-1cm}

\begin{abstract}
 Gliomas, a common type of malignant brain tumor, present significant surgical challenges due to their similarity to healthy tissue. Preoperative Magnetic Resonance Imaging (MRI) images are often ineffective during surgery due to factors such as brain shift, which alters the position of brain structures and tumors. This makes real-time intraoperative MRI (ioMRI) crucial, as it provides updated imaging that accounts for these shifts, ensuring more accurate tumor localization and safer resections. This paper presents a deep learning pipeline combining You Only Look Once Version 8 (YOLOv8) and Segment Anything Model Vision Transformer-base (SAM ViT-b) to enhance glioma detection and segmentation during ioMRI. Our model was trained using the Brain Tumor Segmentation 2021 (BraTS 2021) dataset, which includes standard magnetic resonance imaging (MRI) images, and noise-augmented MRI images that simulate ioMRI images. Noised MRI images are harder for a deep learning pipeline to segment, but they are more representative of surgical conditions. Achieving a Dice Similarity Coefficient (DICE) score of 0.79, our model performs comparably to state-of-the-art segmentation models tested on noiseless data. This performance demonstrates the model's potential to assist surgeons in maximizing tumor resection and improving surgical outcomes.
\end{abstract}

\section{Introduction}

Gliomas are a common type of cancerous brain tumors that account for about 30\% of all brain tumors and 80\% of all malignant brain tumors \cite{zeng2015glioma}. Standard treatment modalities for gliomas include surgery, chemotherapy, and radiation therapy, with surgery often being the preferred option for most neurosurgeons \cite{goldbrunner2018treatment}. The primary goal of surgery is to physically remove as much of the tumor as possible in a process known as resection \cite{wykes2021importance}, in which imaging technologies play a crucial role. Preoperative imaging, particularly MRI, is essential for diagnosing and planning the surgical approach. 

However, the intraoperative success of glioma resection is frequently challenged by several factors. Brain shift, a phenomenon that occurs when the brain changes position during surgery, significantly hinders a surgeon’s ability to accurately locate and resect the tumor \cite{gerard2017brain, brain_shift_golby_lab}. Another complication arises when gliomas infiltrate surrounding brain tissue, making it difficult to clearly delineate tumor margins\cite{whitfield2014imaging, VanHese2022}. As a result, there runs a risk of leaving behind residual tumor cells, which can lead to the recurrence of the glioma or removing too much healthy tissue.

To address these challenges, neurosurgeons have adopted real-time imaging techniques using intraoperative magnetic resonance imaging (ioMRI), which has emerged as the preferred imaging tool for brain tumor operations \cite{rogers2021intraoperative, foroglou2009intra, haydon2013impact, solis2020intraoperative, iMRI_MayoClinic}. ioMRI allows surgeons to update their view of the brain and tumor as the surgery progresses, compensating for brain shift and improving the accuracy of tumor resection. The interpretation of ioMRI images can be time-consuming due to the potential for human error, which can prolong surgery and increase the risk of complications\cite{abernethy2012intra}. Moreover, the process of identifying tumor margins on ioMRI images is manually conducted and subject to human error and variability \cite{shamsuddeen2018assessment}. 

In this paper, we propos a pipeline that utilizes a YOLOv8 model for the detection of gliomas from ioMRI images, followed by the SAM model to refine the segmentation results, thereby ensuring higher accuracy and robustness. To evaluate the robustness of our model, we tested it on augmented MRI images that were simulated through the addition of Gaussian noise to MRI images. These augmented MRI images are similar to ioMRI images, which are generally noisier. Our model achieved a similar dice score to state-of-the-art tumor segmentation models and merits further exploration for use in improving glioma resection outcomes.

\section{Methodology}
\label{headings}
\subsection{Data Preprocessing}
Our model was trained on the open access BraTS 2021 dataset, which is a collection of clinically acquired MR images of annotated glioma tumors from consenting patiends\cite{baid2021rsna,menze2014multimodal,bakas2017advancing,bakas2017segmentation,bakas2017segmentationb}. As the YOLO model can only process colorized images, an image processing function was developed to colorize the grayscale images. This was conducted by assigning an RGB value to each pixel based off its intensity. Another function was developed to create bounding boxes from the ground truth segmentation of the images. Following this function, all images and masks were resized to 256x256 pixels. Finally, the model was trained on both standard MR images and ioMR images that were synthesized through the addition of Gaussian noise. In order to accomplish a dataset of usable ioMR images, the signal-to-noise ratio (SNR) of the BraTS dataset was decreased to mimic ioMR images. ioMR images typically have an SNR of 25 under standard clinical conditions \cite{fei2002automatic}. By decreasing the SNR and resolution, these modified images simulate the qualities of an ioMR image, demonstrated in Figure 1.

\begin{figure}[H]
    \centering
    \includegraphics[width=0.5\linewidth]{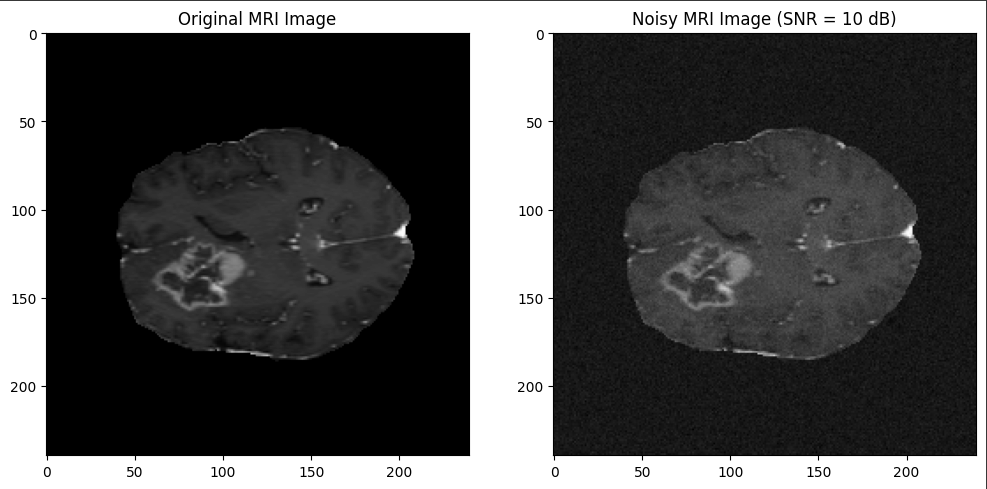}
    \caption{Left: regular MRI image. Right: augmented MRI image with SNR of 10}
    \label{fig:enter-label}
\end{figure}
\subsection{Architecture}
\begin{figure}[H]
    \centering
    \includegraphics[width=0.75\linewidth]{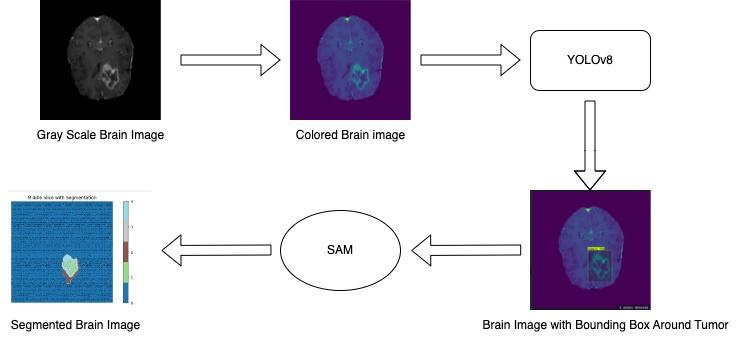}
    \caption{YOLO + SAM architecture; grayscale images are processed through an RGB assignment function, then passed through YOLO in which a bounding box is located around the tumor, the middle coordinate of the bounding box is passed into SAM, finally SAM produces a segmented brain image.}
    \label{fig:pipeline}
\end{figure}

The architecture of the model integrates two state-of-the-art algorithms, YOLO and SAM, to effectively detect and segment glioma tumors. The processed images are first fed into the YOLO model. The YOLO model then identifies the tumor and places a bounding box around it, additionally returning the middle coordinate of the bounding box. Following this general tumor detection, SAM is then used to precisely outline and segment the tumor based on the coordinates provided by YOLO. Figure \ref{fig:pipeline} details this process.
A pre-trained YOLOv8 model was chosen for our application to quickly and accurately detect the approximate location of tumors. YOLOv8 outperforms contemporary models and previous versions of the YOLO algorithm in speed and accuracy \cite{Jocher2023}, making it highly suitable for real-time object detection tasks. Once the MRI images of the brain are passed into the YOLO model, it processes these images through a convolutional neural network (CNN), which extracts essential features and predicts bounding boxes around potential tumors. The YOLO model outputs the center coordinates of the predicted bounding box, which is then passed into SAM as a prompt.

The purpose of the SAM model in the pipeline is to refine the detection results provided by YOLOv8, ensuring that the tumors are accurately and precisely segmented for further analysis. The SAM ViT-b model was selected due to its lightweight nature, allowing for our model to be cost efficient while still maintaining high accuracy. Once the center coordinates of the YOLO bounding box are passed as a prompt into the SAM model, these inputs are used to perform precise segmentation, delineating the exact boundaries of the tumors. The SAM model then produces a detailed probability mask that delineates the tumor regions within the MRI images.

\subsection{Training} 
The model was trained using the BraTS 2021 dataset using both standard MR images and the simulated ioMR images shown in Figure 1. In order to ensure consistency during training, middle slices from the axial plane (slices taken parallel to the X-axis) were extracted from the dataset by selecting the 78th slice of 155 from each image, thus converting the images from 3D to 2D. The middle slice is where the tumor is largest, making it the best choice for training. The specific MRI scan used was T1CE due to its tumor clarity within YOLO.

The YOLO model did not require any training on the BraTS dataset as it was already pre-trained on it. To fine-tune SAM, the middle coordinates of every bounding box in the training set, produced by YOLO, are fed as an initial prompt. SAM was trained on this data over 10 epochs. After the SAM model was finished being trained on the regular BraTS images, YOLO and SAM were then trained on the augmented version, or simulated ioMRI version, of the BraTS images. 

\section{Results}
The proposed YOLO + SAM model was evaluated on an augmented version of the BraTS 2021 dataset. The model was evaluated using a Dice Similarity Coefficient (DICE) score, which is the similarity between two sets of data, in this case, predicted segmentation and ground truth, on a 0 -1 range with 1 indicating perfect overlap. The numerical value is calculated by 2 times the overlap area divided by the total area. The model achieved a DICE score of 0.79 on the augmented BraTS testing set for enhancing tumor (ET), which indicates a strong agreement between the predicted and ground truth segmentation. When compared to other state-of-the-art baseline models, YOLO + SAM has a comparable performance despite running on intentionally noised data. These models include $E_1$ $D_3$ U-Net, Extended VAT method, and NVAUTO; created by Bukhari et al., Peiris et al., and Siddiquee et al. respectively, which were chosen as baselines as they are the state of the art trained on the BraTS 2021 dataset\cite{bukhari2021e1d3}\cite{peiris2021reciprocal}\cite{rahman2021redundancy}. Their models achieved DICE scores of 0.826, 0.814, and 0.86 for ET. The inference times for these models are significantly higher, with estimates of 4 to 8 minutes, 3 to 6 minutes, and 45 to 90 seconds respectively, compared to 15 to 25 seconds for YOLO + SAM. This comparison shows the strong capability of the YOLO + SAM model as it achieved comparable performance to models that tested on images that were noiseless, while YOLO + SAM was tested on images that had extreme amounts of noise. The significantly lower inference time makes YOLO + SAM more suitable for real-world iMRI applications, providing faster and reliable results during surgery.
\vspace{-10pt}  

\begin{figure}[H]
    \centering
    \begin{subfigure}[b]{0.45\textwidth}
        \centering
        \begin{tabular}{|l|c|}
            \hline
            & Dice Score\\ 
            \hline
            \textit{E}$_1$\textit{D}$_3$ U-Net & 0.826\\
            Extended VAT & 0.814\\
            NVAUTO & 0.860\\
            \textbf{YOLO + SAM} & \textbf{0.790}\\
            \hline
        \end{tabular}
        \caption{Comparison of model performances on the BraTS 2021 dataset. Note that, unlike the other models, YOLO+SAM achieved this score on \textit{noised} data, demonstrating striking robustness.}
        \label{tab:model_performance}
    \end{subfigure}
    \hspace{0.02\textwidth}
    \begin{subfigure}[b]{0.4\textwidth}
        \centering
        \includegraphics[width=\textwidth]{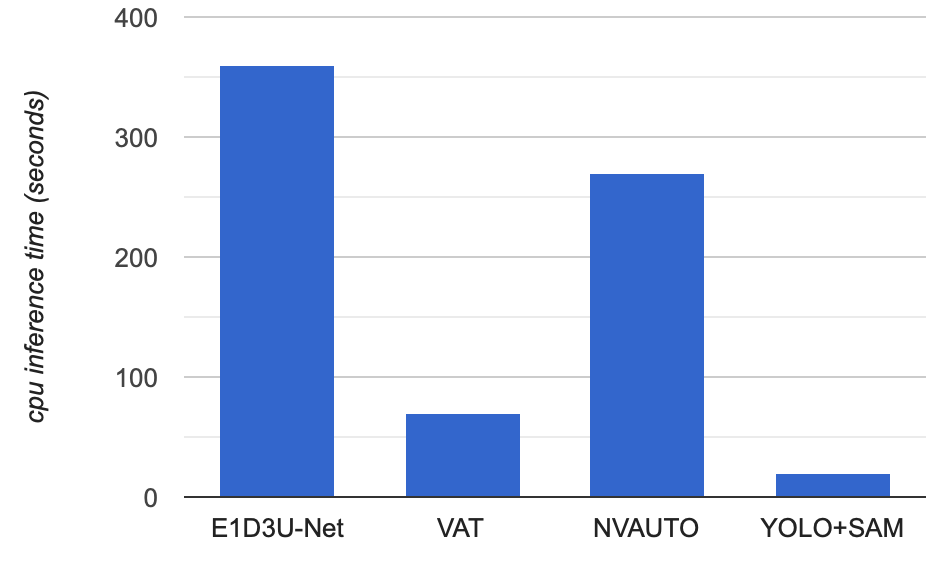}
        \caption{Tumor Segmentation Model Inference Times. Note that, again, unlike the other models, YOLO+SAM achieved this score on \textit{noised} data}
        \label{fig:inference_times}
    \end{subfigure}
    \caption{DICE performance comparison (left) and inference times of various models (right).}
    \label{fig:combined}
\end{figure}
\vspace{-0.7cm}  
\section{Discussion}
Physicians have used computed tomography (CT) scans, positron emission tomography (PET) scans, and MRI to detect and diagnose gliomas in patients \cite{glioma_website}. Historically, machine learning applications for glioma imaging have focused on classification, diagnosis, and preoperative planning. For instance, Hua et al. implemented a cascaded V-Net model ensembling on segmented gliomas, which achieved high accuracy in delineating the whole tumor, tumor core, and enhanced tumor regions on the BraTS 2018 online validation set \cite{hua2020segmenting}. Another study by Shen et al. explores the use of a convolutional neural network combined with near-infrared II (NIR-II) fluorescence imaging, which achieves high sensitivity and specificity in the classification of tumor versus non-tumor intraoperatively \cite{shen2021real}. The YOLO algorithm for object detection was then implemented by Abdulsalomov et al, who developed a YOLOv7 model for the detection of glioma tumors using MRI images, achieving 99.5\% accuracy \cite{abdusalomov2023brain}.

While the aforementioned models report high performances, they cannot be used intraoperatively and do not provide real-time imaging critical for glioma resection. The FL-CNN model proposed does have intraoperative capabilities, but it can only be used on fluorescent images, rendering it infeasible for ioMRI applications. This further clarifies the need for an ioMRI-specific model.

Recent research has shown an abundance of high-resolution, preoperative MRI data, prompting efforts to leverage this data as a proxy for ioMRI. Fei et al. addressed this by simulating low-field interventional MRI images to align real-time interventional MRI images with high-resolution MRI images \cite{fei2002automatic}. By adding noise and creating thicker slices, they successfully simulated 3D images that matched the signal-to-noise ratio of interventional MRI images \cite{fei2002automatic}. Given that interventional MRI and ioMRI have the same fundamental qualities, their method can be used to simulate the dataset necessary to train an effective model \cite{blanco2005interventional}.

In this context, we introduce a novel method using the YOLO algorithm combined with SAM to identify and detect glioma tumors in real time during ioMRI. In this study, we introduced a novel YOLO + SAM model capable of detecting and segmenting glioma tumors using ioMRI images. The model achieved a DICE score of 0.79 for ET and inference time of 15 to 25 seconds, which displays a robust ability for efficient and effective tumor segmentation. This can have a profound impact in the field of glioma surgery as integrating this model with an ioMRI machine could result in improved patient outcomes and more successful surgeries.

To address the limitations of this model, several areas for improvement have been identified. The first is that the YOLO + SAM model we produced was trained solely on simulated ioMRI images, and for future research a model trained on proper clinical ioMRI images could have better performance and accuracy. The second is that the SAM model used for this currently only supports 2D inputs, which according to Zhang et al. could "result in a loss of context information", so an application of SAM to 3D data could be a promising venture \cite{zhang2023segment}. A possible method for incorporating 3D data with SAM is by using TomoSAM which is a 3D slicer extension that uses SAM to help with the segmentation of 3D data from tomography or other imaging methods \cite{semeraro2023tomosam}.

\section{Acknowledgements}
All source code and the text of this paper were authored by Samir Kassam, Angelo Markham, Katie Vo, and Yashas Revanakara, who designed the project following an extensive literature review. We extend our gratitude to Mike Lam and Kevin Zhu for their contributions through lectures on machine learning and research skills, suggested readings, high-level guidance, and constructive comments on the manuscript.

\medskip

\small

\bibliographystyle{abbrvnat}

\bibliography{Gliomas} 

\end{document}